\documentclass[final,5p,times,twocolumn]{elsarticle}

\usepackage[hidelinks]{hyperref}
\usepackage{siunitx}

\usepackage{graphicx}
\usepackage{amsmath}   

\graphicspath{figs}

\begin{document}

\begin{frontmatter}

\title{Characterization of the RD50-MPW4 HV-CMOS pixel sensor}

\author[1]{B. Pilsl\corref{cor1}}
\ead{Bernhard.Pilsl@oeaw.ac.at}

\author[1]{T. Bergauer}
\author[5]{R. Casanova}
\author[1]{H. Handerkas}
\author[1]{C. Irmler}
\author[3]{U. Kraemer}
\author[6]{R. Marco-Hernandez}
\author[6]{J. Mazorra de Cos}
\author[7]{F. R. Palomo}
\author[2]{S. Powell}
\author[1]{P. Sieberer\fnref{fn2}}
\author[3]{J. Sonneveld}
\author[1]{H. Steininger}
\author[2]{E. Vilella}
\author[2]{B. Wade}
\author[2]{C. Zhang}
\author[4]{S. Zhang}

 \cortext[cor1]{Corresponding author}
 \fntext[fn2]{Now at Paul Scherrer Institut (PSI).}

\affiliation[1]{organization={Austrian Academy of Sciences, Institute of High Energy Physics}, 
                 addressline={Nikolsdorfergasse 18},
                 postcode={1050}, 
                 city={Vienna}, 
                 country={Austria}}                 
\affiliation[2]{organization={Department of Physics, University of Liverpool},
                 addressline={Oliver Lodge Building, Oxford Street},
                 postcode={L69 7ZE}, 
                 city={Liverpool},
                 country={UK}}
\affiliation[3]{organization={NIKHEF},
                 addressline={Science Park 105},
                 postcode={1098 XG},
                 city={Amsterdam},
                 country={Netherlands}}
\affiliation[4]{organization={Physikalisches Institut, Rheinische Friedrich-Wilhelms-Universitaet Bonn},
                 addressline={Nussallee 12},
                 postcode={53115}, 
                 city={Bonn},
                 country={Germany}}
\affiliation[5]{organization={Institute for High Energy Physics (IFAE), 
Autonomous University of Barcelona (UAB)},
                 addressline={Bellaterra},
                 postcode={08193}, 
                 city={Barcelona},
                 country={Spain}}
\affiliation[7]{organization={Department of Electronic Engineering, University of Sevilla},
                 addressline={Calle San Fernando 4},
                 postcode={41092}, 
                 city={Sevilla},
                 country={Spain}}
\affiliation[6]{organization={Instituto de Fisica Corpuscular (IFIC), CSIC-UV},
                 addressline={Parque Cientifico, Catedratico Jose Beltran 2},
                 postcode={46980}, 
                 city={Paterna (Valencia)},
                 country={Spain}}

\begin{abstract}
The RD50-MPW4 is the latest HV-CMOS pixel sensor from the CERN-RD50-CMOS working group, designed to evaluate the HV-CMOS technology in terms of spatial resolution, radiation hardness and timing performance. Fabricated by \emph{LFoundry} using a 150nm process, it features an improved architecture to mitigate crosstalk, which has been an issue with the predecessor RD50-MPW3 \citep{Sieberer_2023, Zhang_2024}, allowing more sensitive threshold settings and full matrix operation. Enhancements include separated power domains for peripheral and in-pixel digital readout, a new backside-biasing process step, and an improved guard ring structure supporting biasing up to \SI{500}{V}, significantly boosting radiation hardness. Laboratory measurements and test beam results presented in this paper show significant improvements over its predecessor regarding noise behavior, spatial resolution, and efficiency.
\end{abstract}


\end{frontmatter}

\section{Characteristics}

The RD50-MPW4 comprises a $64 \times 64$ pixel matrix with a pitch of $62 \times 62 \mu m^2$. The threshold of each pixel can be tuned by a \SI{4}{bit} trimDAC. Whilst all sensors are fabricated using a $\SI{3}{k \Omega cm}$ resistivity substrate, a subset of sensors was backside processed to allow backside biasing and to improve radiation hardness. In this additional process step the sensors were thinned from $\SI{300}{\mu m} \rightarrow \SI{280}{\mu m}$, an additional $\mathrm{p^+}$ layer was implanted and the wafer backside was metallised with titanium and aluminium.

\section{Laboratory Assessments}
\label{sec:lab}

\begin{figure}
\centering
\includegraphics[width=0.48\textwidth]{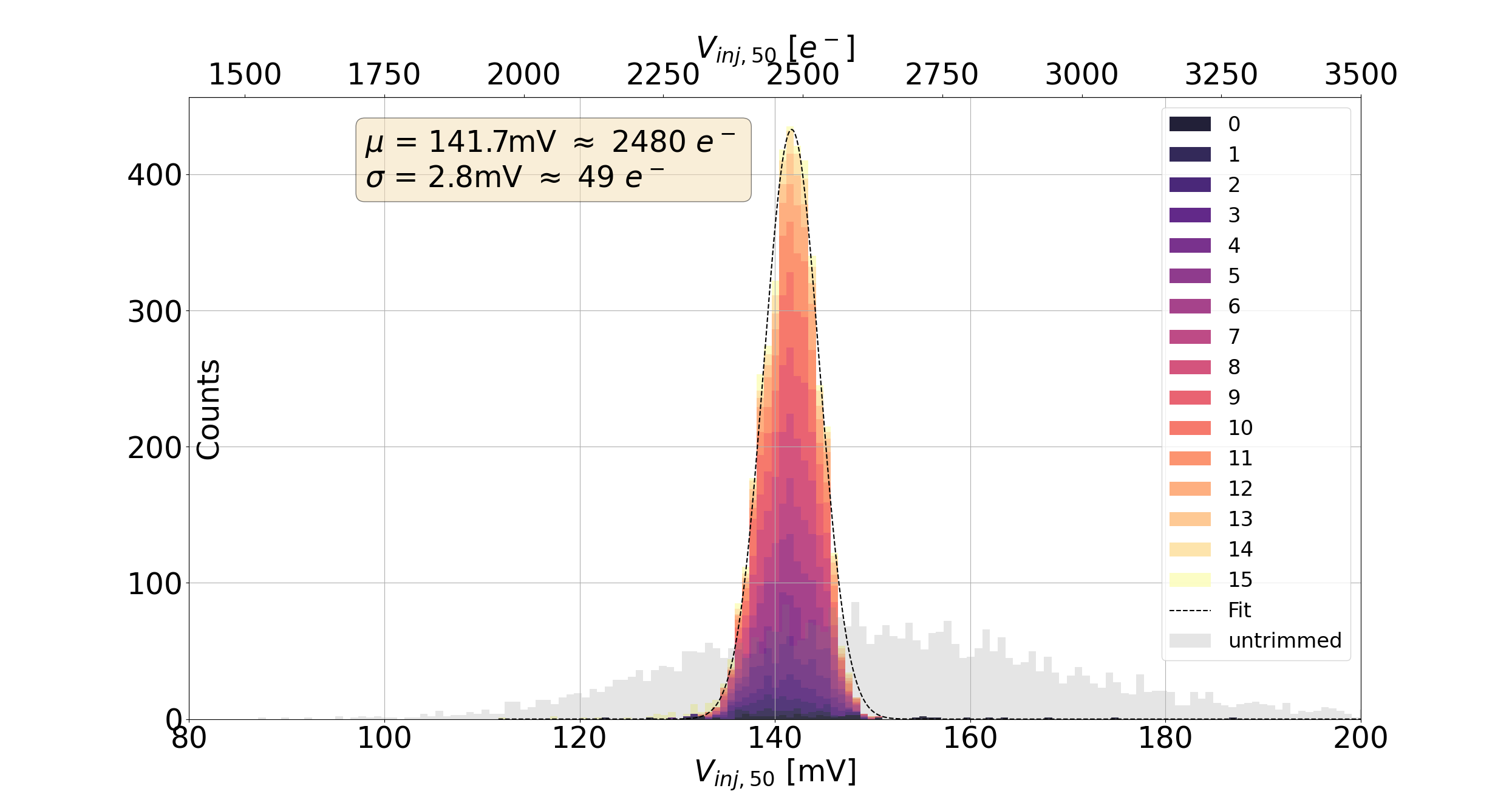}
\caption{Histogram showing $V_{\text{inj, 50}}$ after harmonizing the pixel response by utilizing the trimDACs. The used trimDAC values are color-coded. The grey histogram in the background shows the pixel response with an equal trimDAC value of 7 for all pixels. The global threshold voltage was set to \SI{950}{mV} with the baseline being fixed at \SI{900}{mV}.}
\label{fig:calib}
\end{figure}

To qualify the trimDACs capabilities to harmonize the pixel response, injection measurements with the in-pixel injection capacitance of $C \approx \SI{2.8}{fF}$ (design value) were performed. In this measurement, the threshold voltage was set to a fixed value while the injection voltage was swept on. 100 injections were performed for each injection step, and the digital readout was used to determine the number of detected pulses. The injection voltage $V_{\text{inj, 50}}$ at which 50\% of the injected hits get detected was measured in figure \ref{fig:calib} and shows a standard deviation of $\sigma \approx 50e^-$ after adjusting the in-pixel trimming DACs appropriately. Without trimming those for the individual pixels, a width of $\sigma \approx \SI{330}{e^-}$ is measured.

\section{Test-Beam Results}
\label{sec:testbeam}

To evaluate the sensor's performance in an actual particle beam, a test campaign was conducted at \emph{DESY} \citep{DIENER2019265}. In this campaign, a \SI{4.2}{GeV} electron beam and the \emph{Adenium} telescope \cite{Liu_2023} were used. Analysis of the data was done utilizing the \emph{Corryvreckan} \cite{corry} framework. The standard parameters for bias voltage and threshold settings for both biasing flavors are $V_{\text{Bias}} = \SI{-190}{V}$ and $V_{\text{Thr}} = \SI{30}{mV} \quad \cong \quad Q_{\text{Thr}} \approx \SI{2500}{e^-}$.

\begin{figure}
\centering
\includegraphics[width=0.48\textwidth]{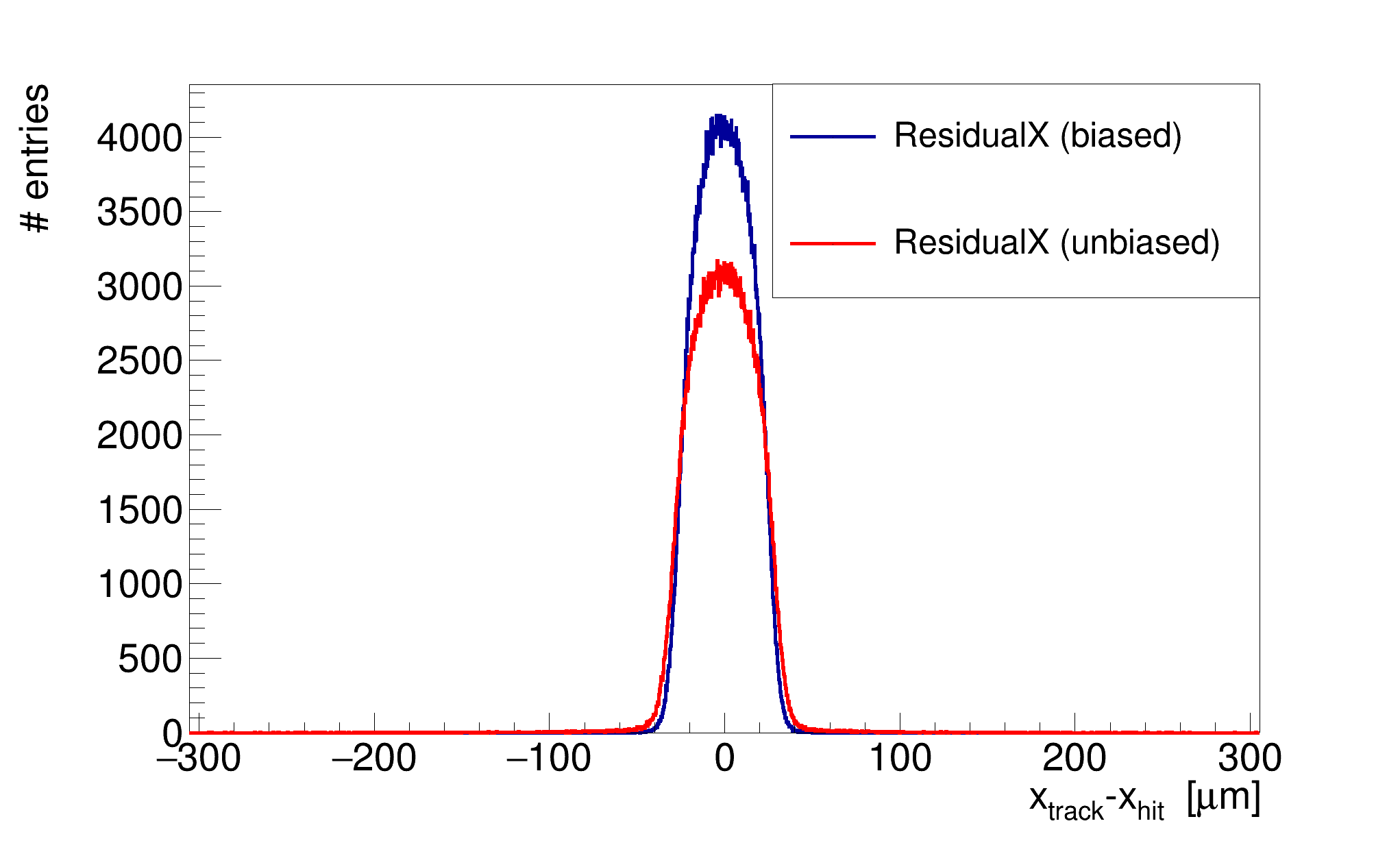}
\caption{The unbiased residuals in the X direction show the difference in the extrapolated track's position at the DUT and the detected cluster center. Evaluating the innermost \SI{99}{\%} for the RMS calculation corresponds to the range of $(\SI{-43.88}{\mu m}, \SI{43.63}{\mu m})$.}
\label{fig:residuals}
\end{figure}

The spatial residuals, as shown in figure \ref{fig:residuals}, allow to depict the spatial resolution of the sensor. By truncating both distributions to the innermost 99\% and calculating the geometric mean of the RMS of the biased and unbiased residuals, the spatial resolution is evaluated as $\SI{15.79}{\mu m}$ in X and $\SI{15.39}{\mu m}$ in Y, with the results presented in table \ref{tab:res}.   The binary resolution of $\SI{62}{\mu m} / \sqrt{12} \approx \SI{17.9}{\mu m}$ is exceeded by taking advantage of the average cluster size of 1.3 pixels per cluster and applying a charge weighted center of gravity calculation for the cluster positions. The minor discrepancy between the two dimensions still must be clarified and is under investigation.

\begin{table}
\begin{tabular}{c|c|c}
Characterisitcs& X-Axis & Y-Axis\\ \hline
RMS (unbiased Residuals) &  $\SI{16.99}{\mu m}$ & $\SI{17.35}{\mu m}$ \\
RMS (biased Residuals) & $\SI{14.67}{\mu m}$ & $\SI{13.65}{\mu m}$  \\ 
Geometric mean & $\SI{15.79}{\mu m}$ & $\SI{15.39}{\mu m}$
\end{tabular}
\caption{Spatial resolution characteristics extracted from the analysis of the spatial residuals.}
\label{tab:res}
\end{table}

\begin{figure}[htbp]
\centering
\includegraphics[width=0.48\textwidth]{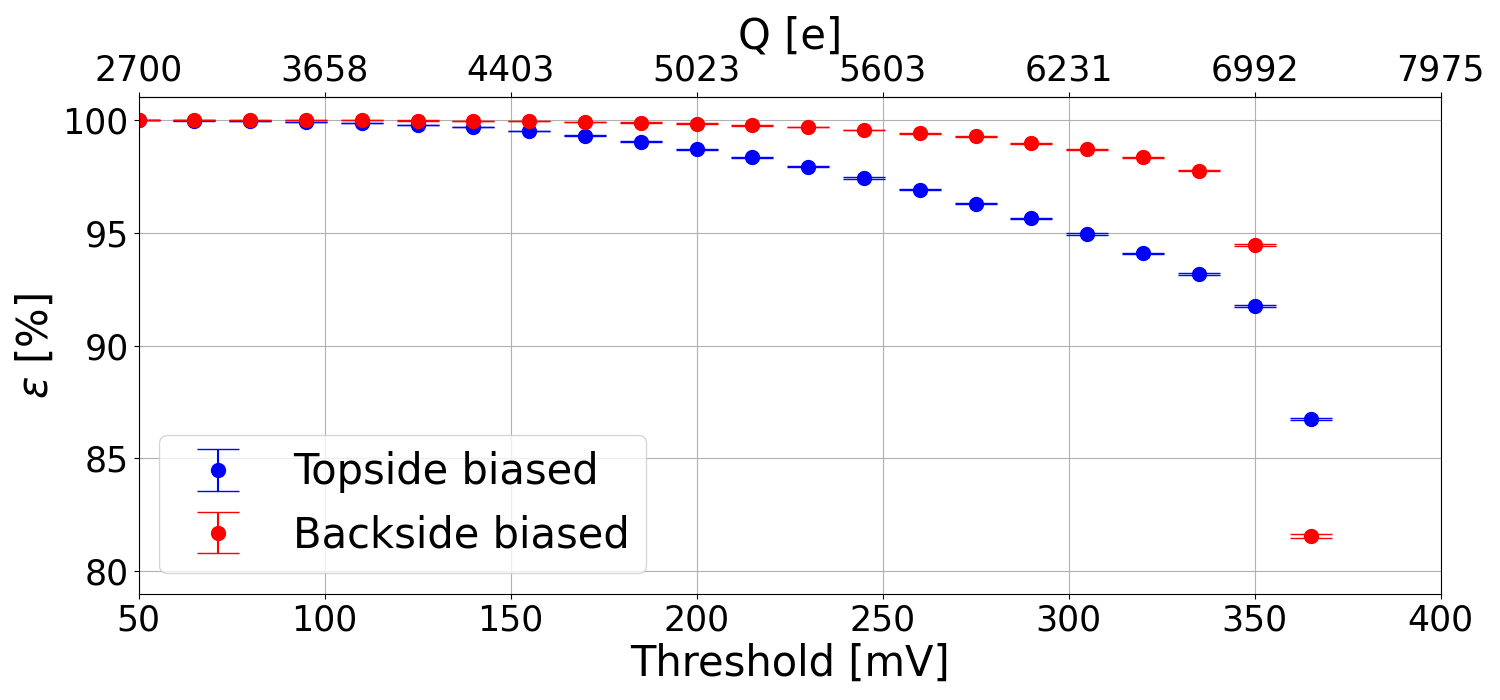}
\caption{The total efficiency as a function of the applied threshold.}
\label{fig:effiVsThr}
\end{figure}

In figure \ref{fig:effiVsThr} efficiency measurements show full efficiencies $>99\%$ up to thresholds of \SI{200}{mV} which corresponds to $\mathcal{O}(\SI{5000}{e^-})$. The efficiency drop at higher thresholds is caused by a degraded in-pixel-efficiency at the pixel corners. These effects are more pronounced in the topside-biased samples than the backside-biased ones. The electric field of the backside biasing approach, which is expected to be more uniform, is suspected to be the main cause for the better performance. Further measurements, especially with irradiated samples, are planned to clarify the differences.

\section{Conclusion}
\label{sec:conclusion}
Laboratory and test beam measurements have demonstrated that the RD50-MPW4 sensor successfully addressed several issues present in its predecessor. A comparison between the topside and backside biasing schemes reveals that backside biasing is superior, as evidenced by efficiency at high thresholds.

\section{Acknowledgements}
\label{sec:ack}
This work has been partly performed in the framework of the CERN-RD50 collaboration. 
The measurements leading to these results have partially been performed at the Test Beam Facility at DESY Hamburg (Germany), a member of the  Helmholtz Association (HGF). The research leading to these results has received funding from the European Union's Horizon Europe research and innovation program under grant agreement no. 101057511.

\bibliographystyle{elsarticle-num-names} 
\bibliography{bibfile}

\end{document}